\documentclass[letterpaper]{article}
\usepackage[top=3cm,bottom=3cm,left=3cm,right=3cm,marginparwidth=1.75cm]{geometry}
\usepackage{listings}
\usepackage{xcolor}
\usepackage{multirow}
\usepackage{dcolumn}
\usepackage[english]{babel}
\usepackage[utf8x]{inputenc}
\usepackage[T1]{fontenc}
\usepackage{palatino}
\usepackage{amsmath}
\usepackage{afterpage}
\usepackage{graphicx, subcaption}
\usepackage[colorinlistoftodos]{todonotes}
\usepackage[colorlinks=true, linkcolor=blue, citecolor=blue]{hyperref}

\usepackage{booktabs}
\usepackage{gensymb}

\newcommand{\be}{\begin{equation}}  
\newcommand{\ee}{\end{equation}}  
\newcommand{\bea}{\begin{eqnarray}}  
\newcommand{\eea}{\end{eqnarray}}  
\begin{document}

\vspace*{1.2cm}

\thispagestyle{empty}
\begin{center}

{\LARGE \bf Toward the Frontiers of Particle Physics With the Muon~$g\textrm{-}2$ Experiment\footnote{Fermilab report FERMILAB-CONF-20-463-E.}}

\par\vspace*{7mm}\par

{

\bigskip

\large \bf Eremey Valetov\footnote{ORCID: 0000-0003-4341-0379.
Presently affiliated with Tsung-Dao Lee Institute of Shanghai Jiao Tong University, with Michigan State University, and with Lepton Dynamics, LLC.}

On Behalf of the Muon~$g\textrm{-}2$ (E989) Collaboration at Fermilab

}

\bigskip

{\large \bf  e-mail: evaletov@fnal.gov}

\bigskip

{Lancaster University, the Cockcroft Institute, and Michigan State University}

\bigskip

{\it Presented at the 3rd World Summit on Exploring the Dark Side of the Universe \\Guadeloupe Islands, March 9-13 2020}

\end{center}



\begin{abstract}
The Muon~$g\textrm{-}2$ Experiment (E989) at Fermilab has a goal of measuring the muon anomaly ($a_\mu$) with unprecedented precision using positive muons. This measurement is motivated by the difference between the previous Brookhaven $a_\mu$ measurement and Standard Model prediction exceeding three standard deviations, which hints at the possibility of physics beyond the Standard Model. Muons are circulated in a storage ring, and the measurement requires a precise determination of the muon anomalous precession frequency (spin precession relative to momentum) from the resulting decay positron time and energy measurements collected with calorimeters. The average magnetic field seen by the muons needs to be known with high precision, and so the storage ring magnetic field is shimmed to be very uniform and is continually monitored with nuclear magnetic resonance (NMR) probes. Detailed Muon Campus beamline and muon storage ring simulations are also required for quantifying beam dynamics and spin-related systematic effects in the determination of the muon anomalous precession frequency, e.g. muon losses during the measurement window. At the time of the conference, the experiment has recently commenced Run-3\footnote{Data-taking in long-running experiments is often divided into campaigns called ``runs''. One run in the Muon~$g\textrm{-}2$ Experiment roughly corresponds to one year.}, and the release of Run-1 physics results is planned for 2020.
\end{abstract}
  
\section{Introduction}
\label{S:intro}

The Muon~$g\textrm{-}2$ Experiment (E989), located at the Muon Campus of Fermilab, measures the  muon magnetic anomaly $a_\mu$ using antimuons $\mu^+$ (``muons'' for brevity) circulating around a storage ring with a highly uniform magnetic field within $\pm0.5\%$ of the momentum $p_{0}=3.094\:\mathrm{GeV}/c$. This momentum $p_{0}$, called the ``magic'' momentum, is defined by $p_{0}=m/\sqrt{a_\mu}$, where $m$ is the muon mass, and which makes the spin precession relative to the muon momentum independent of any external transverse electric fields (see also Eq.~\ref{eq:omega-a}).

The current Standard Model prediction of $a_\mu$ is \cite{arXiv:2006.04822}
\be
a_{\mu}^{\textrm{SM}}=\left(116\,591\,810\pm43\right)\times10^{-11},
\ee
while the current experimental world average is \cite{Patrigiani}
\be
a_\mu^{\textrm{Exp}}=\left(116\,592\,089\pm63\right)\times10^{-11}.
\ee
The contributions to $a_\mu$ in the Standard Model (SM) are quantum electrodynamical (QED), electroweak, leading order (LO) hadronic (Had), hadronic light-by-light (LbL), and higher order (HO) hadronic. Ref. \cite{Jegerlehner_2018} shows recent values of these SM contributions to $a_\mu$.

Recent progress on the computation of $a_\mu^{\rm SM}$ includes improvements in the dispersive evaluations of the hadronic vacuum polarization (VP) contributions, which rely on experimental measurements of the hadronic cross section. The recent estimate of the leading order hadronic VP contribution from the Muon~$g\textrm{-}2$ Theory Initiative, $a_{\mu}^{\textrm{Had, LO VP}}=\left(693.1\pm4.0\right)\times10^{-10}$~\cite{arXiv:2006.04822}, represents a combination of direct energy scan results from CMD-3, SND, and KEDR experiments and radiative return results from BABAR, KLOE, and BESIII experiments.

A recent achievement in the theory regarding $a_\mu$ is the calculation \cite{Guelpers:2020Um} of hadronic vacuum polarization $a_{\mu}^{\textrm{Had, VP}}$ and hadronic light-by-light $a_{\mu}^{\textrm{Had, LbL}}$ contributions from first principles using lattice QCD. Several collaborations are working on this, including RBC/UKQCD and Mainz. The precision of this calculation is subject to improvement, and the calculation is a work-in-progress.

In case the Muon~$g\textrm{-}2$ Experiment yields a beyond--Standard Model value of $a_\mu$, some of the possible contributions to the discrepancy could be dark photons, inelastic dark matter (iDM), supersymmetry (SUSY), extra dimensions, axion-like particles \cite{PhysRevD.94.115033}, and additional Higgs bosons \cite{Iguro_2019}. 

The Muon~$g\textrm{-}2$ Experiment (E989) at Fermilab improves upon its predecessor, the Muon~$g\textrm{-}2$ Experiment (E821) at Brookhaven National Laboratory (BNL), by using a higher intensity muon beam, having an improved muon storage function, employing better beam dynamics modeling, having higher field uniformity and better field monitoring, and achieving reduced spin precession frequency systematics. The technical design projection \cite{E989TDR} of E989 is to obtain ${\sim}20$ times more data and a ${\sim}3$-fold reduction of systematic errors compared to E821.

\section{Systems and Methods of the Muon~$g\textrm{-}2$ Experiment}
\label{S:1}

The Muon~$g\textrm{-}2$ storage ring uses a toroidal C-magnet of $7\:\mathrm{m}$ radius with $1.45\:\mathrm{T}$ magnetic field. Electrostatic quadrupoles provide vertical beam focusing. Muons are injected into the storage ring through an inflector, which cancels the $1.45\:\mathrm{T}$ field at injection, and are deflected onto the design orbit by three kickers located $90\degree$ downstream of the inflector.

By tuning the momentum of muons in the Muon~$g\textrm{-}2$ storage ring to the "magic" momentum $p_{0}=3.094\:\mathrm{GeV}/c$, the anomalous precession frequency is proportional to both the parameter of interest $a_{\mu}$ and the traversed magnetic field:
\be
\label{eq:omega-a}
\vec{\omega}_{a}=-a_{\mu}\frac{q_\mu\vec{B}}{m_\mu},
\ee
where $\omega_a$ is the muon anomalous spin precession frequency, $q_\mu$ is the muon charge, $m_\mu$ is the muon mass, and $B$ is the magnetic field, which must be highly uniform for the measurement. Thus, in principle, to measure $a_{\mu}$ it is necessary to measure ${\omega}_{a}$ and $B$ with high precision.

In E989, we determine the magnetic field in the storage region by measuring at-rest proton spin precession with nuclear magnetic resonance (NMR) probes. The formula to calculate $a_{\mu}$ by the experiment is as follows:
\be
a_{\mu}=\frac{g_{e}}{2}\frac{m_{\mu}}{m_{e}}\frac{\omega_{a}}{\tilde\omega_p^\prime\left(T\right)}\frac{\mu^{\prime}_{p}\left(T\right)}{\mu_{e}},
\ee
where $g_e$ is the $g$-factor of the electron, $m_e$ is the electron mass, $\omega_p$ is the angular precession frequency of the proton, $\mu_e$ and $\mu_p$ are the spin magnetic moments of the electron and the proton respectively, the prime symbol (`$\prime$') denotes shielding of the proton, the tilde on $\omega^{\prime}_p$ denotes the muon-weighted average over space and time, and $T$ is the temperature. The frequencies $\omega_a$ and $\omega_p$ are obtained from decay positron time spectra and NMR, respectively.

The values of $g_e$, $m_\mu/m_e$, and $\mu_e/\mu_p$ are known from CODATA \cite{RevModPhys.88.035009} with uncertainties that are low enough to not be a concern. The proposed systematic and statistical errors on $\omega_a$, according to the technical design report (TDR) \cite{E989TDR} of the experiment,  are $70\:\mathrm{ppb}$ and $100\:\mathrm{ppb}$, respectively. The proposed systematic error of $\omega_p$ as per the TDR is $70\:\mathrm{ppb}$.

The detector systems of the Muon~$g\textrm{-}2$ storage ring for measuring $\omega_a$ are straw trackers and calorimeters. The straw trackers reconstruct decay positron trajectories, while the calorimeters detect decay positron energies and arrival times.

\begin{figure}[ht!]
\begin{center}
\includegraphics[width=0.76\textwidth]{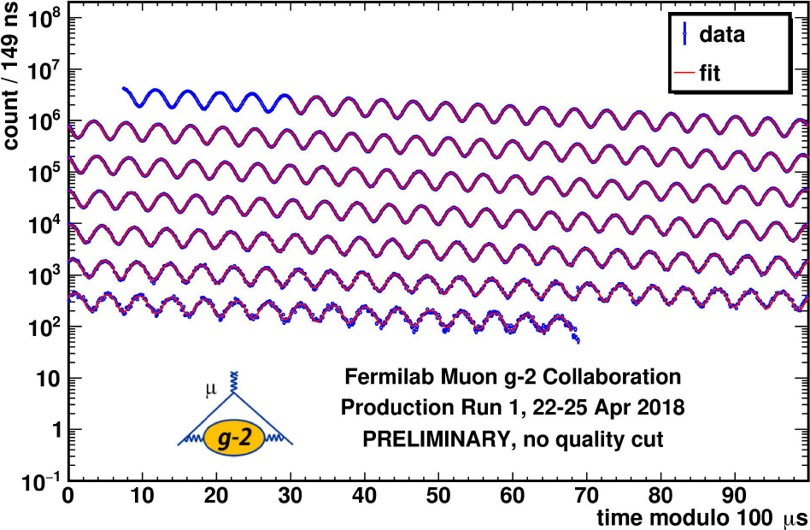}
\end{center}
\caption{\label{fig:wiggle-plot} A decay positron time distribution (``wiggle plot'', preliminary data).}
\end{figure}

The decay positron time distribution, or the ``wiggle plot'' (see Fig.~\ref{fig:wiggle-plot} for an example), has the same count oscillation frequency as $\omega_a$. This oscillation occurs because the decay positron energy spectrum has maxima and minima when the muon spin and momentum are aligned and anti-aligned, respectively.

The basic fitting function for the wiggle plot is 
\be
\label{eq:ff}
f\left(t\right)=N_{0}\exp\left(-\lambda t\right)\left[1+A\cos\left(\omega_{a}t+\phi\right)\right],
\ee
where $N_0$ is an initial positron count and $\lambda$ is the exponential decay constant. Fitting functions with a higher number of parameters are usually used, such as a 22-parameter fitting function.

From Eq.~\ref{eq:ff}, we can see that any time-dependence of $\phi$ due to additional time-varying phenomena in the ring affects the measured value of $\omega_a$. For example, if $\phi=\phi\left(t\right)=\phi_0+\phi_{1}t$, then
\be
\begin{alignedat}{1}\cos\left(\omega_{a}t+\phi\right) & =\cos\left(\omega_{a}t+\phi_{0}+\phi_{1}t\right)=\\
 & =\cos\left(\left(\omega_{a}+\phi_{1}\right)t+\phi_{0}\right).
\end{alignedat}
\ee 
Such effects are referred to as `early-to-late' phase change systematics and need to be carefully studied and quantified. One such study is briefly discussed in Sec. 3.

Passive shimming is performed by inserting tiny metal pieces to increase the magnetic field. Using passive shimming, the magnetic field was made three times more uniform than in E821 at BNL. Active shimming is also used.

For measuring $\omega_p$, a mobile device called the trolley is used to sweep the storage ring with 17 NMR probes. There are also 378 NMR probes installed outside the vacuum chamber at fixed locations. Scanning of $\omega_p$ using the trolley is performed regularly, where the trolley moves around the storage ring inside the vacuum chamber while the beam is off. The fixed NMR probes monitor $\omega_p$ between the trolley runs.

Tables with expected uncertainties on $\omega_a$ and $\omega_p$ are available in the TDR \cite{E989TDR} of the experiment. We will  discuss some details regarding the lost muons systematic error of $\omega_a$ in the next section.

\section{End-to-End Beamline Simulations and Systematic Analyses}
\label{S:2}

In this section, we will focus on the recent contributions to the Muon~$g\textrm{-}2$ experiment by the author of this proceedings paper.

There are \emph{G4Beamline} \cite{g4bl} and \emph{MARS} \cite{Mokhov:2017klc,Mokhov:2014soa} beam dynamics models \cite{PhysRevAccelBeams.20.111003} of the target station that include the generation of the secondary beam of $\pi^{+}$ particles by impinging the proton primary beam on the Inconel target, the lithium lens that is used to focus the secondary beam, the copper collimator that provides radiation shielding to the pulsed magnet (PMAG) that is located directly downstream of it, and the PMAG.

We carefully checked the \emph{G4Beamline} and \emph{MARS} models of the target station against its documentation and drawings, and we consulted a number of specialists regarding the layout of the target station and the geometry of its particle optical elements. We revised the \emph{G4Beamline} and \emph{MARS} models based on this verification.

\begin{figure}[ht!]
\begin{center}
\includegraphics[width=1.0\textwidth]{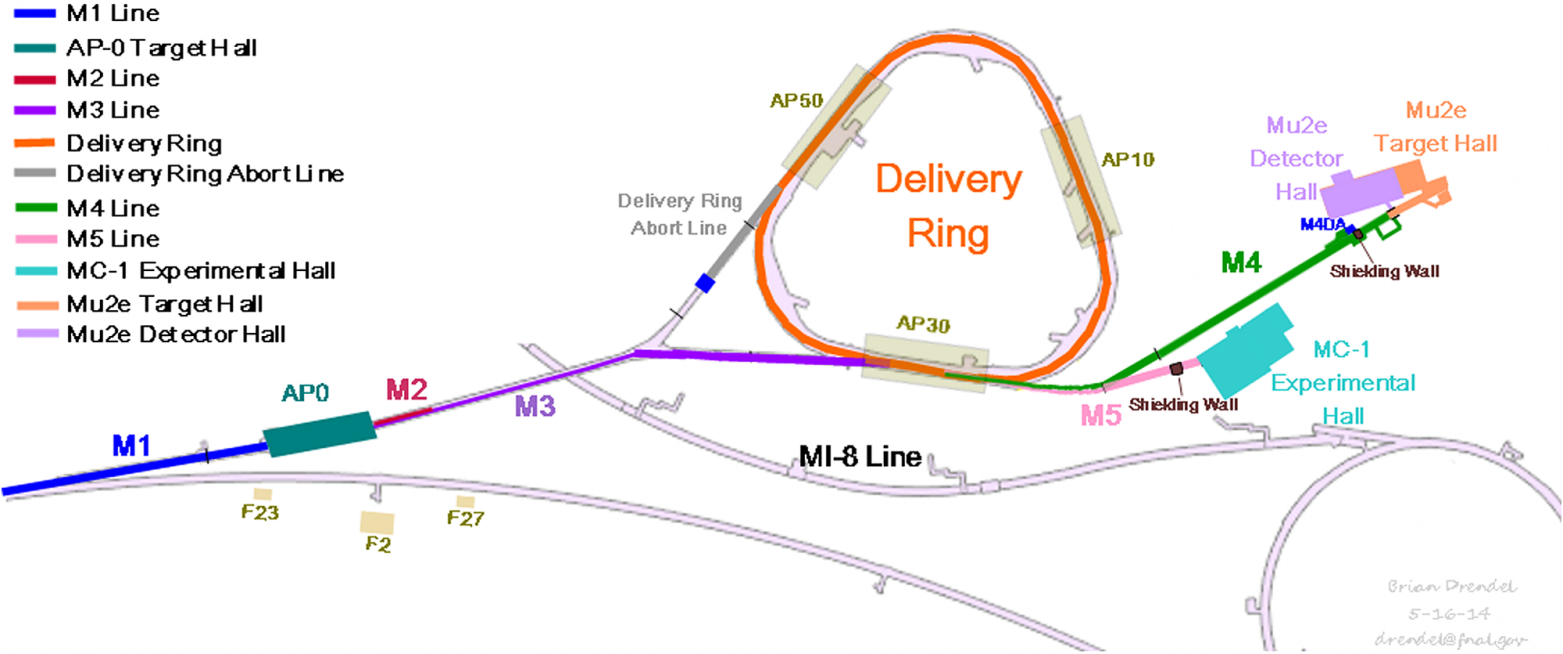}
\end{center}
\caption{\label{fig:muon-campus} Layout of the beamlines of the Muon Campus at Fermilab. (Image used with permission from Brian Drendel.)}
\end{figure}

To accurately model and quantify the early-to-late phase change systematics, we must understand the potential sources of these systematics. To that end, we have performed \cite{GM19979} high-statistics simulations of the beamlines of the Muon Campus of Fermilab, starting from the Muon~$g\textrm{-}2$ target station (AP0), through the M2 and M3 beamlines, for four turns around the Delivery Ring, and through the M4 and M5 beamlines (see Fig.~\ref{fig:muon-campus}). The simulations further continued through the injection channel, including the inflector, and up to $2000$ turns around the Muon~$g\textrm{-}2$ storage ring.

In the experiment, ${\sim}1\times10^{12}$ protons-on-target (PoT) yield several thousand muons after the first turn around the storage ring, considering factors including the tertiary beam nature of the muon beam, the losses of ${\sim}50\%$ on the aperture of the inflector, and further losses of ${\sim}99.5\%$ in the first turn around the storage ring. For end-to-end beamline simulations to yield sufficiently high statistics in the storage ring, it is thus necessary to run simulations of at least ${\sim}1\times10^{12}$ PoT in aggregate.

The high-statistics simulations of the Muon~$g\textrm{-}2$ beamlines were performed with $3\times10^{12}$ PoT using \emph{G4beamline}-3.04 from the upstream end of the target station to the end of the M3 beamline, and using \emph{BMAD} \cite{Sagan:Bmad2006,Korostelev:2016qaf} from the end of the M3 beamline to up to $2000$ turns in the Muon~$g\textrm{-}2$ storage ring. High-performance computing (HPC) resources of the National Energy Research Scientific Computing Center (NERSC) were used for the simulations.

The simulations kept track of the initial phase space coordinates of the muons, enabling the analysis of their correlations with the characteristics of the muons as they circulate around the storage ring.

Our studies based on the simulation results include a $g\textrm{-}2$ phase-versus-momentum-deviation study, which is useful for the determination of the systematic shift of $\omega_a$ due to momentum-dependent muon losses; a muon losses study; and a principal component analysis (PCA) with $k$-means clustering. The purpose of the PCA was to find correlations between orbital and spin coordinates in the storage ring and the position within the beamlines where the muon was originally produced \cite{GM19979}.

\begin{figure}[ht!]
\begin{center}
\includegraphics[width=0.8\textwidth]{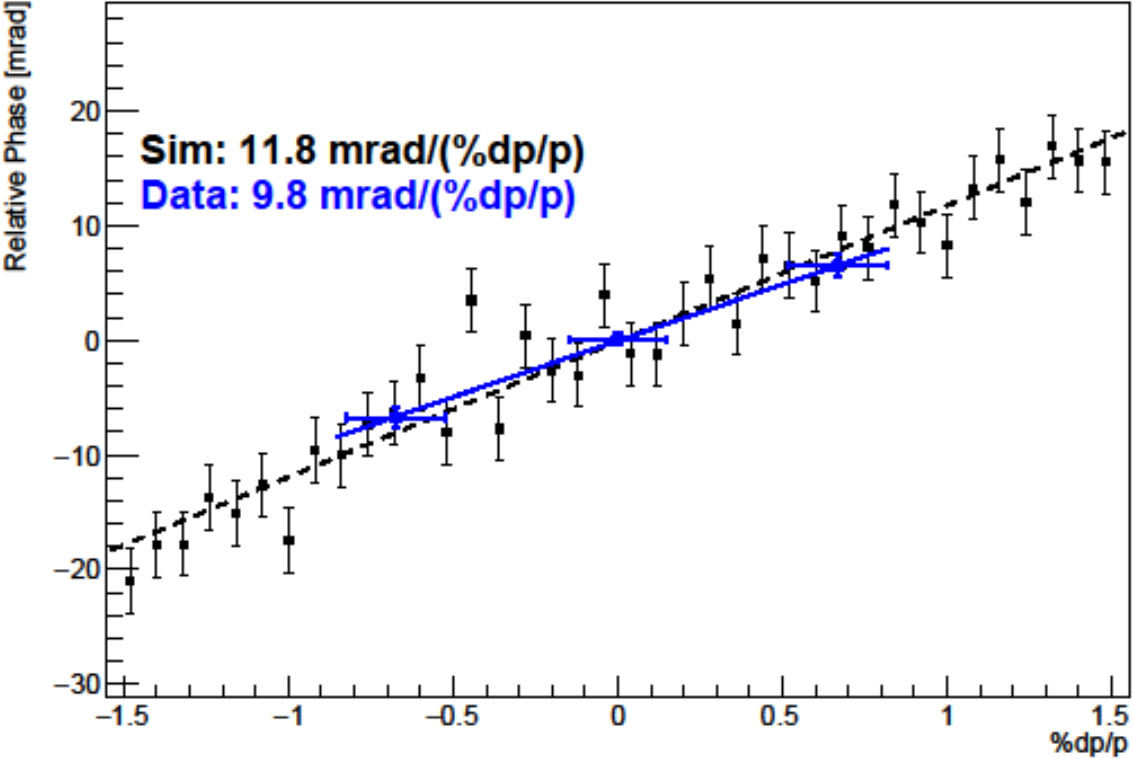}
\end{center}
\caption{\label{fig:phase-momentum-dependence} Momentum dependence of the initial phase $\phi$ in Eq.~\ref{eq:ff} due to the magnetic dipoles in the Delivery Ring. The simulations (black plot markers) are in good agreement with experimental data (blue, preliminary). The experimental data were obtained by studying the behavior of muons with momenta above and below the magic momentum in the storage ring.}
\end{figure}

The $g\textrm{-}2$ phase-versus-momentum-deviation slope at injection from our simulations is in good agreement with experimental data, as Fig.~\ref{fig:phase-momentum-dependence} shows. The equilibrium radius $x_{\mathrm{e}}$ of a muon in the storage ring depends on its momentum deviation $\Delta p/p_{0}$ with a linear approximation $\Delta p/p_{0}=\eta x_{\mathrm{e}}$, where $\eta$ is the periodic dispersion. Because of the amplitude-dependent tune shifts and the scattering on the collimators, the momentum-dependent $x_{\mathrm{e}}$ causes a momentum dependence of the muon losses in the storage ring. Thus, there is an early-to-late spin phase change in the storage ring due to muon losses.

The muon distribution at the end of the M5 beamline from the $3\times10^{12}$ PoT  simulations is being used as a new initial muon distribution for simulations of the Muon~$g\textrm{-}2$ storage ring using \emph{gm2ringsim} \cite{gm2ringsim}, which is a custom simulation code based on \emph{Geant4} and Fermilab's \emph{art} event-processing framework \cite{Green:2012gv}.

While work on the early-to-late phase change due to muon losses is still in progress, we expect a $10-20\:\mathrm{ppb}$ error of $\omega_a$ due to muon losses, which is far below the overall $70\:\mathrm{ppb}$ TDR systematic error on the spin precession, and meets the TDR goal of $20\:\mathrm{ppb}$ for the systematic error due to muon losses.

\section{Conclusion}

As of the time of the conference, the Muon~$g\textrm{-}2$ experiment is running well and has entered Run-3. We are making progress on the analysis, and we are expecting to release the first results in 2020.

We have developed sophisticated modeling tools and data-driven approaches to quantify systematics, such as the muon loss phase.

The author of this paper performed high-statistics simulations of the Muon Campus beamlines using \emph{G4beamline} and \emph{BMAD}, and a number of systematic analyses based on the $3\times10^{12}$ PoT simulations. The momentum dependence of the initial $g\textrm{-}2$ phase from the $3\times10^{12}$ PoT simulations is in good agreement with preliminary experimental data. The muon distribution from the $3\times10^{12}$ PoT simulations is used  for \emph{Geant4}-based \emph{gm2ringsim} simulations of the Muon~$g\textrm{-}2$ storage ring.


\section*{Acknowledgements}

We are thankful to Ian Bailey, Jim Morgan, Diktys Stratakis, Sergei Striganov, Chris Polly, Alex Herrod, Maxim Korostelev, Yun He, Anthony Leveling, Alex Keshavarzi, Ronald Lebeau, Dean Still, Nikolai Mokhov, and Vladimir Tishchenko for productive and interesting discussions. This work was supported by the Cockcroft Institute of Accelerator Science and Technology, a Science and Technology Facilities Council facility, under STFC grant no. ST/P002056/1. This manuscript has been authored by Fermi Research Alliance, LLC under Contract No. DE-AC02-07CH11359 with the U.S. Department of Energy, Office of Science, Office of High Energy Physics. This research used the resources of the National Energy Research Scientific Computing Center (NERSC), a U.S. Department of Energy Office of Science User Facility operated under Contract No. DE-AC02-05CH11231. This material is based upon work supported by the U.S. Department of Energy, Office of Science, under Contract No. DE-FG02-08ER41546 and Contract No. DE-SC0018636.

\bibliographystyle{JHEP} 
\bibliography{references.bib}

\providecommand{\href}[2]{#2}\begingroup\raggedright\begin{thebibliography}{10}

\bibitem{arXiv:2006.04822}
T.~{Aoyama} et~al., {\it {The anomalous magnetic moment of the muon in the
  Standard Model}},  {\em arXiv e-prints} (2020) arXiv:2006.04822.

\bibitem{Patrigiani}
{C. Patrignani et al. (Particle Data Group)}, {\it {Review of Particle
  Physics}},  {\em Chinese Physics C} {\bf 40} (2016), no.~10 100001. See also
  its 2017 update.

\bibitem{Jegerlehner_2018}
F.~Jegerlehner, {\it {The Muon $g\textrm{-}2$ in Progress}},  {\em Acta Physica
  Polonica B} {\bf 49} (2018), no.~6 1157.

\bibitem{Guelpers:2020Um}
V.~Guelpers, {\it {{Recent Developments of Muon $g\textrm{-}2$ from Lattice
  QCD}}},  in {\em Proceedings of 37th International Symposium on Lattice Field
  Theory {\textemdash} PoS(LATTICE2019)}, vol.~363, p.~224, 2020.

\bibitem{PhysRevD.94.115033}
W.~J. Marciano, A.~Masiero, P.~Paradisi, and M.~Passera, {\it {Contributions of
  axionlike particles to lepton dipole moments}},  {\em Phys. Rev. D} {\bf 94}
  (2016) 115033.

\bibitem{Iguro_2019}
S.~Iguro, Y.~Omura, and M.~Takeuchi, {\it {Testing the 2HDM explanation of the
  muon $g\textrm{-}2$ anomaly at the LHC}},  {\em Journal of High Energy
  Physics} {\bf 2019} (2019), no.~11.

\bibitem{E989TDR}
J.~Grange et~al., {\it {Muon ($g$-$2$) Technical Design Report}},  {Design
  Report FERMILAB-FN-0992-E}, Muon $g$-$2$ Collaboration, Fermi National
  Accelerator Laboratory, Batavia, IL, 2015.

\bibitem{RevModPhys.88.035009}
P.~J. Mohr, D.~B. Newell, and B.~N. Taylor, {\it {CODATA recommended values of
  the fundamental physical constants: 2014}},  {\em Rev. Mod. Phys.} {\bf 88}
  (2016) 035009.

\bibitem{g4bl}
{Muons, Inc.}, {\it G4beamline (version 3.04)},  2017.
\newblock Available from http://www.muonsinternal.com/muons3/G4beamline
  [accessed 14-Sep-2020].

\bibitem{Mokhov:2017klc}
N.~V. Mokhov and C.~C. James, {\it {The MARS Code System User's Guide Version
  15(2016)}},  {Technical Report FERMILAB-FN-1058-APC}, Fermi National
  Accelerator Laboratory, Batavia, IL, 2017.

\bibitem{Mokhov:2014soa}
N.~Mokhov, P.~Aarnio, Y.~Eidelman, K.~Gudima, A.~Konobeev, V.~Pronskikh,
  I.~Rakhno, S.~Striganov, and I.~Tropin, {\it {{MARS15 Code Developments
  Driven by the Intensity Frontier Needs}}},  {\em Prog. Nucl. Sci. Tech.} {\bf
  4} (2014) 496--501.

\bibitem{PhysRevAccelBeams.20.111003}
D.~Stratakis, M.~E. Convery, C.~Johnstone, J.~Johnstone, J.~P. Morgan,
  D.~Still, J.~D. Crnkovic, V.~Tishchenko, W.~M. Morse, and M.~J. Syphers, {\it
  {Accelerator performance analysis of the Fermilab Muon Campus}},  {\em Phys.
  Rev. Accel. Beams} {\bf 20} (2017) 111003.

\bibitem{GM19979}
E.~Valetov, {\it {Muon $g$-$2$ End-to-End Beamline Simulations, and Systematic
  Analyses of Muon Losses and Origin Effects}},  {G-2 Experiment Document
  GM2-doc-19979}, Muon $g$-$2$ Collaboration, Fermi National Accelerator
  Laboratory, Batavia, IL, 2020.

\bibitem{Sagan:Bmad2006}
D.~Sagan, {\it {Bmad: A relativistic charged particle simulation library}},
  {\em Nucl. Instrum. Meth.} {\bf A558} (2006), no.~1 356--359. Proceedings of
  the 8th International Computational Accelerator Physics Conference.

\bibitem{Korostelev:2016qaf}
M.~Korostelev, I.~Bailey, A.~Herrod, J.~Morgan, W.~Morse, D.~Stratakis,
  V.~Tishchenko, and A.~Wolski, {\it {{End-to-End Beam Simulations for the New
  Muon G-2 Experiment at Fermilab}}},  in {\em {7th International Particle
  Accelerator Conference}}, 2016.
\newblock Paper WEPMW001.

\bibitem{gm2ringsim}
{Muon $g$-$2$ Collaboration}, {\it gm2ringsim (version 9.52.00)},  2020.
\newblock Available from https://cdcvs.fnal.gov/redmine/projects/gm2ringsim
  [accessed 12-Sep-2020].

\bibitem{Green:2012gv}
C.~Green, J.~Kowalkowski, M.~Paterno, M.~Fischler, L.~Garren, and Q.~Lu, {\it
  {The Art Framework}},  {\em J. Phys. Conf. Ser.} {\bf 396} (2012) 022020.

\end{thebibliography}\endgroup


\end{document}